\documentclass{aastex63}
\pdfoutput=1

\newcommand{\rev}{ }

\revised{2020 Mar 14}
\accepted{2020 Mar 16}
\published{2020 Apr 1}
\submitjournal{ApJL}
\reportnum{ApJ 892, L36}

\begin{document}

\title{The Low Earth Orbit Satellite Population and Impacts of the SpaceX Starlink Constellation}
\shorttitle{Impacts of Starlink}
\shortauthors{McDowell}

\correspondingauthor{Jonathan McDowell}
\email{jcm@cfa.harvard.edu}
\author[0000-0002-7093-295X]{Jonathan C. McDowell}

\affiliation{Center For Astrophysics | Harvard \& Smithsonian \\
60 Garden St, \\
Cambridge, MA 02138, USA}

%\nocollaboration{2}

\begin{abstract}

I discuss the current low Earth orbit artificial satellite population
and  show that the proposed `megaconstellation' of circa 12,000 Starlink
internet satellites would dominate  the lower part of Earth orbit, below
600 km, with a latitude-dependent areal number density of between 0.005 and 0.01 objects per square degree  at airmass
$<2$. Such large, low altitude satellites appear visually bright to
ground observers, and the initial Starlinks are naked eye objects. I
model the expected number of illuminated satellites as a function of
latitude, time of year, and time of night and summarize the range of possible
consequences for ground-based astronomy. In winter at lower latitudes
typical of major observatories, the satellites will not be illuminated
for six hours in the middle of the night. However,
at low elevations near
twilight at intermediate latitudes (45-55 deg, e.g. much of Europe)
hundreds of satellites may be visible at once to naked-eye observers at dark sites.
\end{abstract}


\section{Introduction} \label{sec:intro}

On 2019 May 24, the US company SpaceX launched the first batch of sixty prototype Starlink Internet-distribution satellites
into a 430 km circular Earth orbit. Within hours, amateur and professional astronomers expressed surprise and concern at the
brightness of the slowly dispersing string of satellites, with most reports suggesting a magnitude of $V\sim 1$ to $3$ for each satellite
\citep{King19}. Over the next few days, the satellites faded by several magnitudes as they adjusted the orientation of their solar
panels, and lost a further 0.5 mag as their orbits were raised to their operational height of 550 km, but nevertheless remained naked eye
objects from dark sites.
{\rev In this paper, by `naked-eye' I will mean a visual magnitude of 6 or brighter, which is a reasonable approximation for detectability
in dark skies. Objects of magnitude 4 or brighter are easily noticeable under such conditions. The Starlink satellites, as shown below, typically
lie between these values.
}

Since the early space age, satellite tracks have been a constant but relatively minor annoyance to
ground-based observers.
An early discussion arose from a 1980 COSPAR panel on `Potentially Environmentally
Detrimental Activities in Space' (PEDAS). The PEDAS reports were published in a special issue of
Advances in Space Research; \cite{Eberst82} said
`At present the effect of satellite trails appearing on Schmidt plates
is more of a nuisance than a problem', with a total of 4643 tracked and
mostly faint objects in orbit at the time. However, F. Graham \cite{Smith82}, shortly before
becoming the Astronomer Royal,
concluded `the cumulative effect of an increasing
number of long-lived satellites represents a very serious hazard [to optical observation].'
That prediction is now coming true with the prospect of tens of thousands  of
orbiting objects bright enough to be visible to the naked eye.

 A first attempt to simulate these effects for the first phase
of the Starlink constellation was presented by \cite{Seitzer20}. In this paper I
present similar calculations for the currently approved full
constellation, and describe the current demographics of the low Earth
orbit (LEO) satellite population, which Starlink is starting to transform and
dominate.

Several other companies in the United States, Europe and China have proposed similar `megaconstellations' of thousands to tens of thousands
of satellites. {\rev Many of these are for telecommunications, but Earth imaging systems have also been suggested.
Of the near-term proposals, Starlink is the one } with the greatest expected light-pollution impact and is the first to be extensively deployed.

The Starlink satellites are 260 kg in mass
and consist of a flat panel about 3 metres across on which communications antennae and
propulsion systems are mounted, together with a solar panel at right
angles to the main bus that is about 9 metres long. At
zenith, at the 550 km altitude of the initial constellation layer, they
will therefore subtend an angle of between $1\arcsec$ and $4\arcsec$ depending
on orientation.

\section{The Low Earth Orbit satellite population}

Low Earth Orbit (LEO) is generally considered to extend from around
80--100 km (below which satellites cannot remain in orbit, \cite{McDowell18}) 
to about 2000
km (above which the intensity of the trapped radiation belts make it
more difficult for satellites to operate,  \cite{IADC07}). To avoid
collisional runaway \citep{Kessler78}, current recommendations advocate that space objects
be removed from LEO within 25 years of the end of operations
\citep{IADC07}. Below about 600 km, the effects of atmospheric drag will
ensure satellite reentry on this timescale for most satellites, without
the need for any special action.

As an indirect result of these different lifetimes, the
satellite populations in lower LEO (100--600 km) and upper LEO (600--2000
km) are qualitatively different.

\begin{figure}[ht!]
\plotone{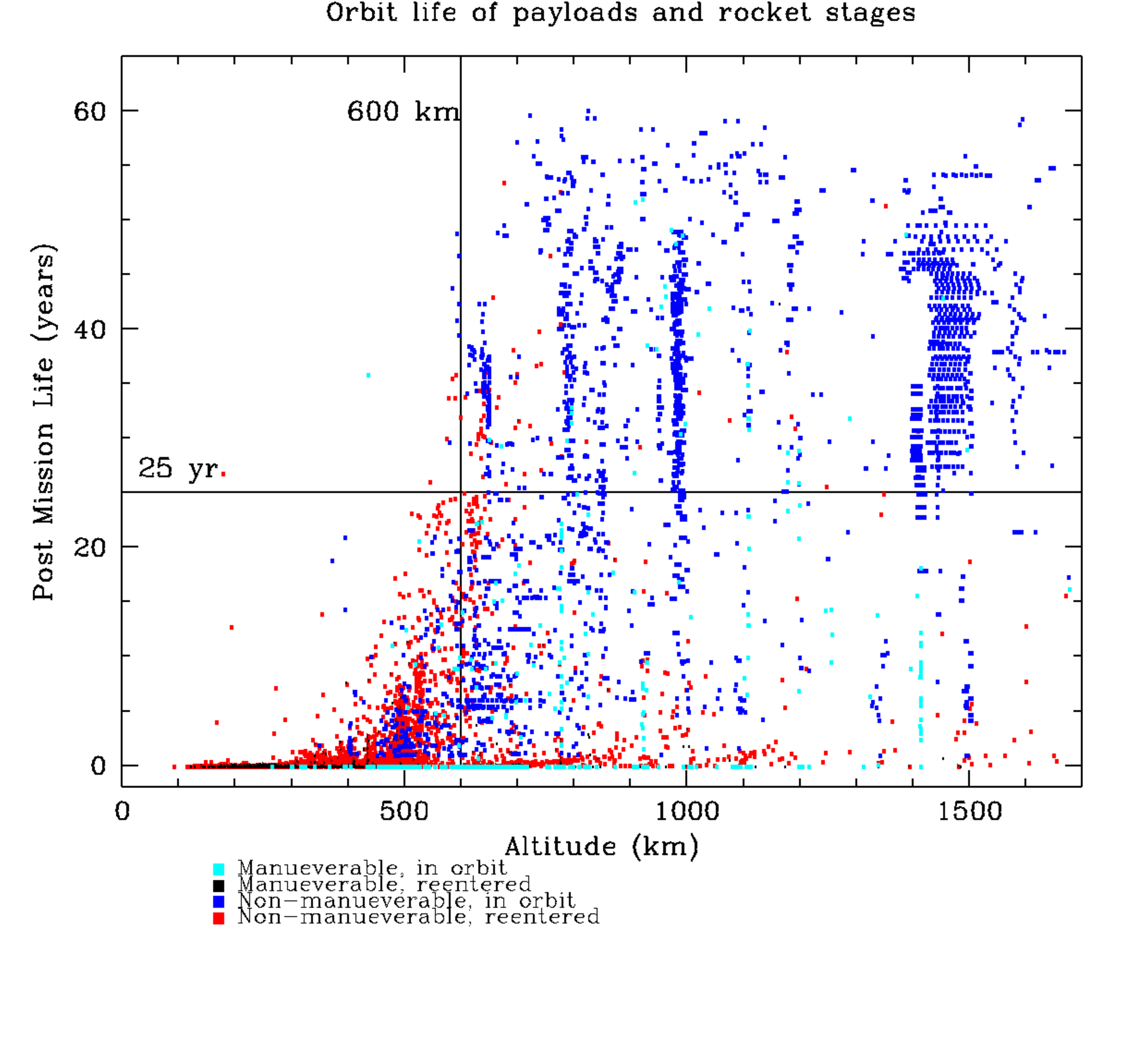}
\caption{Post mission orbital lifetime for tracked payloads and rocket stages versus average altitude of
initial operational orbit.  \label{fig:life}
Maneuvering satellites are plotted in cyan (objects still in orbit)
and black (objects which have reentered; almost all at the bottom left of the
plot) while non-maneuvering satellites are plotted in
blue (objects still in orbit) and red (objects which have reentered).}
\end{figure}

I have analyzed the US catalog of space objects \citep{USSF20}, extracted orbital
data for all tracked objects, and categorized each object as described below.
The 600 km dichotomy can be seen in Figure \ref{fig:life}, where I plot the post-mission life
of payloads and rocket stages in LEO from 1957 to the beginning of 2020, generated
from a database of Earth satellites I maintain. The post-mission life
is calculated from the end of active operations of the object until its reentry
or until the present. 
The active life of payloads was determined by review of mission documents and public sources
as well as analysis of orbital maneuvers.
Maneuvering satellites are plotted in cyan (objects still in orbit)
and magenta (object which have reentered) while non-maneuvering satellites are plotted in
blue (objects still in orbit) and red (objects which have reentered). It can be seen that
the upper left quadrant defined by the black lines
(objects in orbits below 600 km which are more than 25 years past their end of mission) is almost empty.

%  test/mac rf2b
\clearpage

\begin{figure}[ht!]
\includegraphics[width=7.5in]{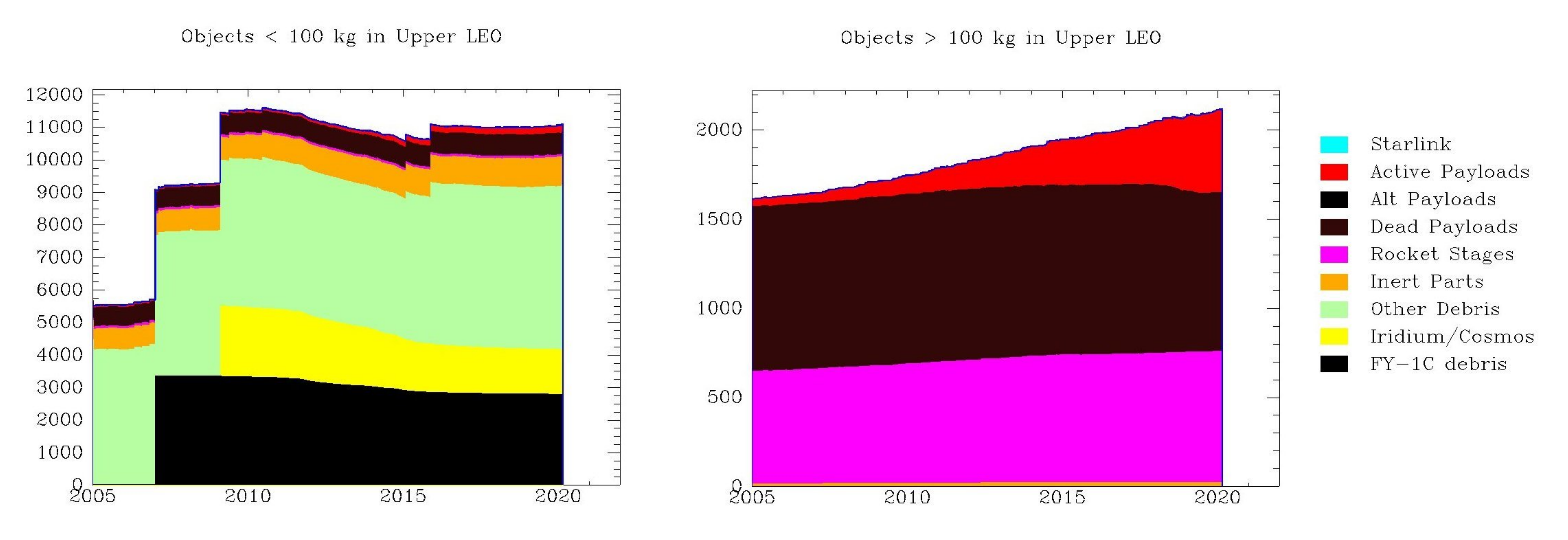}
\caption{Evolution of tracked artificial Earth satellite population in the 2005-2020
period, in upper LEO (600 to 2000 km). Left: Small ($<$100 kg) satellites.
Right: Large ($>$100 kg) satellites.
The small object population is dominated by debris objects; large objects are mostly
dead payloads and rocket stages. Evolution shows steady growth. \label{fig:o1}
}
\end{figure}

\begin{figure}[ht!]
\includegraphics[width=7.8in]{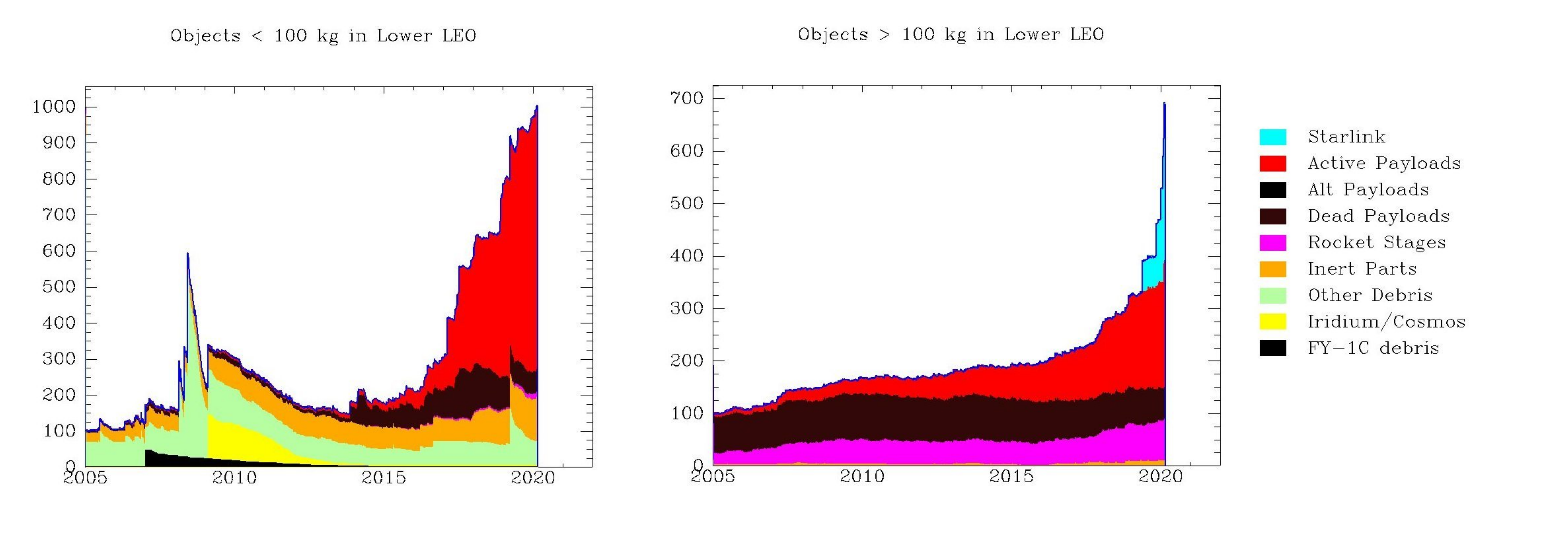}
\caption{Evolution of tracked artificial Earth satellite population in the 2005-2020
period, in lower LEO (200 to 600 km). Left: Small ($<$100 kg) satellites.
Right: Large ($>$100 kg) satellites, including Starlink satellites. 
Small object population shows a rapid increase of small active payloads beginning
in 2016 (the `cubesat revolution'). Large object population shows a recent spike as
Starlink satellites begin to dominate in 2020. Objects contributing to the large
object population in this figure are likely
to be visible to the naked-eye.  \label{fig:o2}}
\end{figure}

In Figs.~\ref{fig:o1} and \ref{fig:o2} I illustrate the evolution of the LEO
artificial satellite population between 2005 and 2020, separately for upper and lower LEO
and for big (mass above 100 kg) and small (below 100 kg) objects. The population is
divided into nine categories, color-coded in the figures as noted: 
(1) Starlink satellites (cyan; right panel in Fig \ref{fig:o2}).;
(2) other active payloads (red),
(3) `alternative' payloads, or `special cases' (a small category of objects for which payload versus
component status is arguable; brown; 
(4) dead (no longer operating) payloads (black);
(5) discarded rocket stages (magenta); 
(6) inert parts or components, 
such as launch vehicle adapters, optics covers, and despin systems (orange); 
(7) debris from
accidents such as collisions, propellant system explosions or battery explosions, or 
from deliberate events such as space weapons tests, excepting the next two categories (light green);
(8) debris from the Chinese space weapons test of 2007 (black); 
(9) debris from the accidental
Iridium-Cosmos collision of 2009 (yellow).

In Table~\ref{table:stats} I summarize the state of the orbital population on 2020 March 1.

\begin{table}[h]
\caption{Estimated status of the LEO satellite population on 2020 March 1. \label{table:stats}} 

\begin{tabular}{p{1.7in}rrrr}
\hline
                  & \multicolumn{2}{c}{Upper LEO ($>600$ km)}  & \multicolumn{2}{c}{Lower LEO ($<600$ km)} \\              
                  &  Small   & Big  &   Small  & Big  \\
                  &   ($<100$ kg)  & ($>100$ kg) &   ($<100$ kg)  & ($>100$ kg) \\
\hline
Starlink           &    0 & 0 & 0 & 299\\ 
Active payloads (excluding Starlink)& 229  & 465 & 731 & 243 \\
Special cases     &     2 &  3 & 0 & 0\\
Dead payloads     &   667 & 887 & 60 & 58\\
Rocket stages     &  62   & 734 & 16 & 78\\
Inert parts       & 899    & 24 & 124 & 10\\
Debris (general)  &  5041 & 2 & 62 & 0\\
Debris (2009 collision)& 1382 & 0 & 3 & 0 \\
Debris (2007 test)&  2801 & 0 &2 & 0\\
\hline
Total              &  11083 & 2115 & 998 & 688\\
\hline
\end{tabular}

\begin{center}
{ Note. The division between
lower and upper LEO is set at 600 km; the division between big and small satellites is set at 100 kg.
Note the significant contribution of Starlink satellites to the big/low category.}
\end{center}

\end{table}

Figure \ref{fig:o1} shows that the evolution of the population in the higher
part of LEO is relatively gradual. The small object population is dominated
by debris objects, and the main changes are sudden increases caused by
individual debris events such as the 2007 Chinese test and the 2009 collision.
The large object population is dominated by dead payloads and discarded rocket
stages. In contrast, the lower part of LEO characterized in Figure \ref{fig:o2}
is quite different. The total number of objects is lower, and since 2016 the small object population has been dominated
by a rapidly increasing number of active payloads -- the cubesats. The large
object population was similar to higher altitudes, but since 2019
shows a dramatic increase in the population due to the Starlink launches.

On 2019 May 27, after the first Starlink launch, SpaceX founder Elon
Musk noted \footnote{\label{Musk19} https://twitter.com/elonmusk/status/1132897322457636864}
that `there are already 4900 satellites in
orbit ... Starlink won't be seen by anyone unless looking very carefully
and will have $\sim 0\%$ impact on advancements in astronomy.'  This is
a rather misleading statement, as satellites are normally only naked-eye
if they are  both large and in the lowest orbits, i.e. those in the right hand panel
in Fig \ref{fig:o2}. As the figure shows, fewer than 400 objects were in
that category prior to the first Starlink launch.  Starlink satellites
are already in the majority in this category following the sixth launch
which took place on 2020 March 18. In the near future, with 1584 Starlinks in the
550 km orbit in the initial constellation and ultimately as many as 9000
in that regime with all proposed deployments (as discussed in the next
section), Starlink will completely dominate the naked-eye object
population by factors of 4 to 20.

% space/sdb/test mac rf2 macro for orbit life  via cmd/mvl
% space/sdb/nstats for fig2

\clearpage

\clearpage

\section{Model of the Starlink constellation}

The initial constellation of 1584 satellites approved by the US Federal Communications Commission \citep{SpaceX16} and modelled
by \cite{Seitzer20} has been superseded. Further FCC filings extend the proposed constellation to about 12,000 satellites \citep{SpaceX17}, \citep{SpaceX19}.
Filings under the USASAT-NGSO-3 label with the International Telecommunications Union (ITU) in 2019 
October suggest as many as 30,000 satellites are
envisaged, but in this paper I will restrict myself to the 12,000 satellite case. The filed orbital properties of the constellation
are summarized in Table \ref{table:constell}, where I group the eight shells specified by SpaceX into
three similar-altitude layers useful for summary analysis. Only layer A 
satellites are currently being deployed, with proposed layers B and C
expected several years from now. Satellites in layers A and B will be grouped
in separate orbital planes distinguished by their longitude of ascending node.

\begin{table}[h]
\caption{Orbital altitudes and inclinations of proposed constellation. \label{table:constell}}

\begin{tabular}{llllll}
\hline
Layer  &  Shell & Altitude & Inclination & Planes & Satellites \\ 
       &        & (km)     &  (deg)      &         & \\
\hline
A      &   1    & 550       & 53.0        & 72   & 1584\\
B      &   2    & 1110      & 53.8        & 32   & 1600\\
B      &   3    & 1130      & 74.0        & 8    &  400\\
B      &   4    & 1275      & 81.0        & 5    &  375\\
B      &   5    & 1325      & 70.0        & 6    &  450 \\
C      &   6    &  346      & 53.0        &      & 2547 \\
C      &   7    &  341      & 48.0        &      & 2478 \\
C      &   8    &  336      & 42.0        &      & 2493 \\
\hline
\end{tabular}

\begin{center}
{ Note: Total
satellites 11927 (1584 in layer A, 2825 in layer B and 7518 in layer C).
Layer C satellites are not grouped in orbital planes. Total satellites below 600 km
(layers A and C only) will therefore be 9102.  }
\end{center}
\end{table}

\begin{figure}[ht!]
\plotone{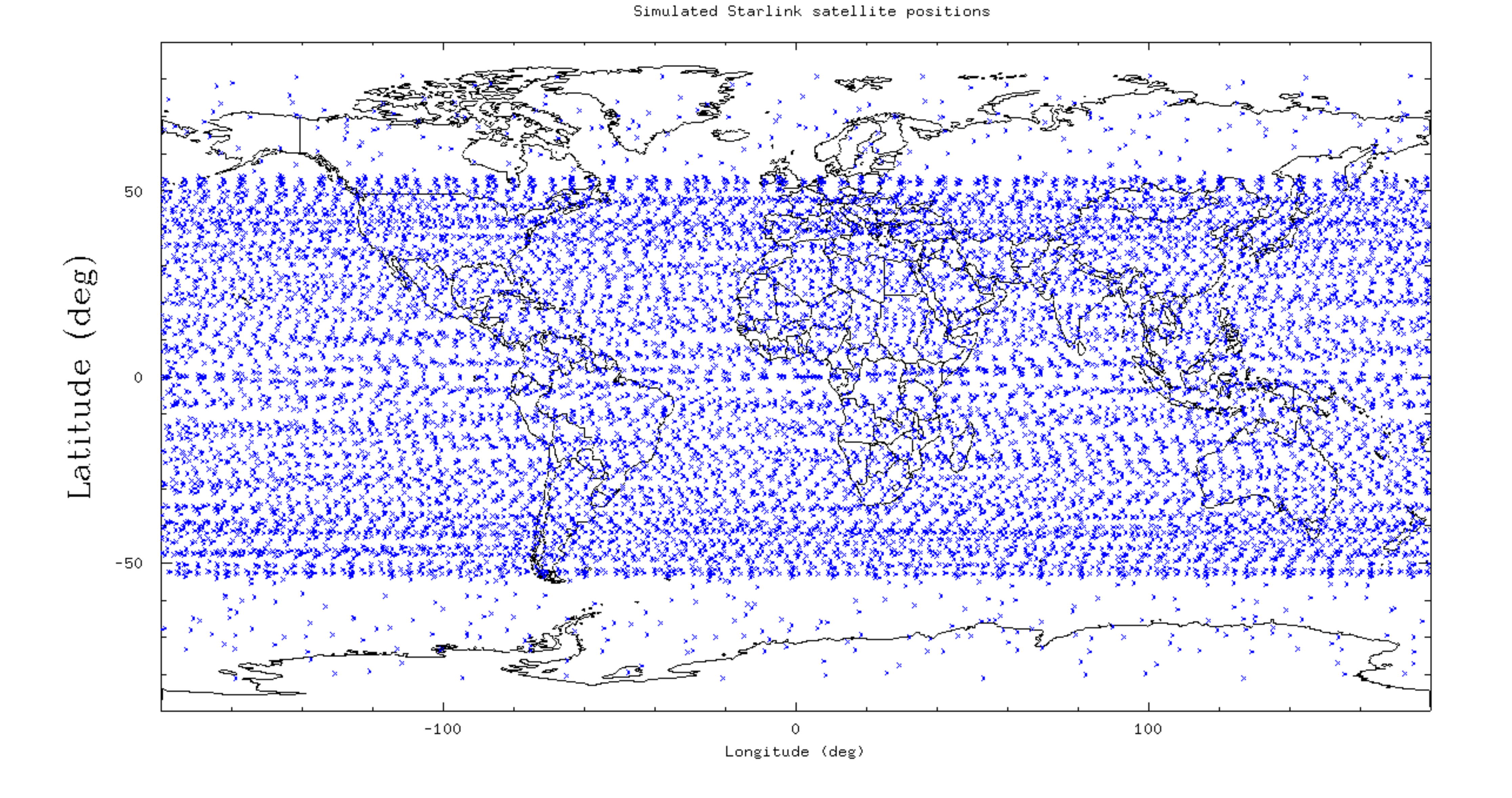}
\caption{Simulated instantaneous distribution of Starlink satellites. The constellation is dense up to about 53 degrees latitude
with a smaller number of satellites at higher latitudes. \label{fig:s2}}
\end{figure}

\begin{figure}[ht!]
\includegraphics[height=4.0in]{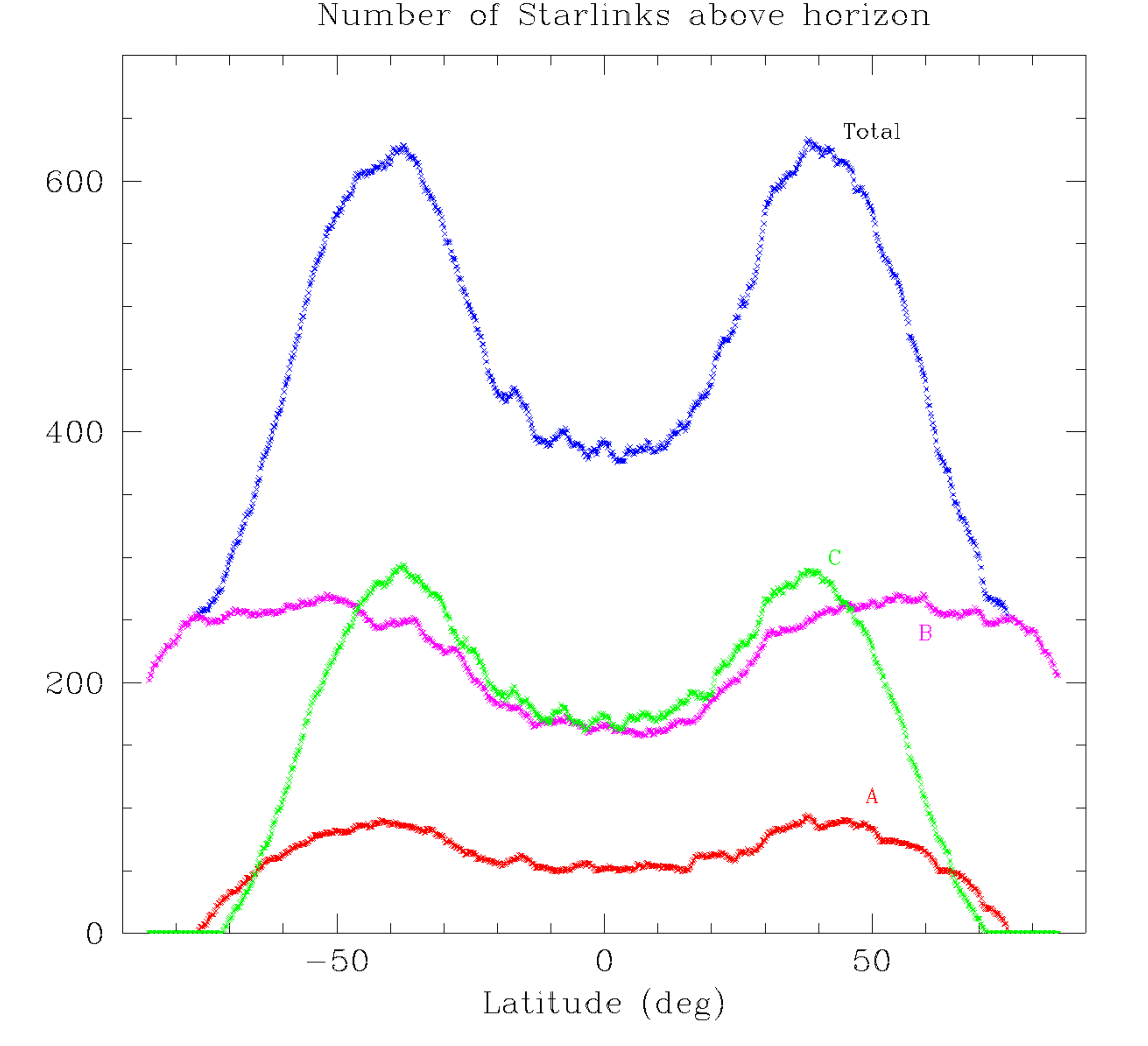}
\caption{Simulated number of Starlink satellites above horizon (but not necessarily illuminated) as a function of latitude \label{fig:s1}. Separate contributions
from layers A, B and C are also shown. The finite number of satellites causes the irregularities in the curves. 
The figure illustrates that the number of satellites is largest at latitudes near the
orbital inclination of the bulk of the satellites and smallest at the equator.
\label{fig:lat}}
\end{figure}

I simulated the instantaneous state of the constellation at a typical time (Fig. \ref{fig:s2})
assuming the satellites in a given orbital plane are equally spaced around the orbit,
and calculated
the number above the horizon as a function of latitude (Fig. \ref{fig:lat}). Since the bulk of the constellation
is in orbits with inclination to the equator of around $48^{\degr} - 54^{\degr}$, the instantaneous distribution peaks at
those latitudes. 

I then simulate the number of these satellites that are illuminated by
the Sun as a function of date and time of night. I do not attempt to model the
reflectivity of the satellite as a function of phase, but simply count which satellites
are in view of the Sun at a given time.

\begin{figure}[ht!]
\plottwo{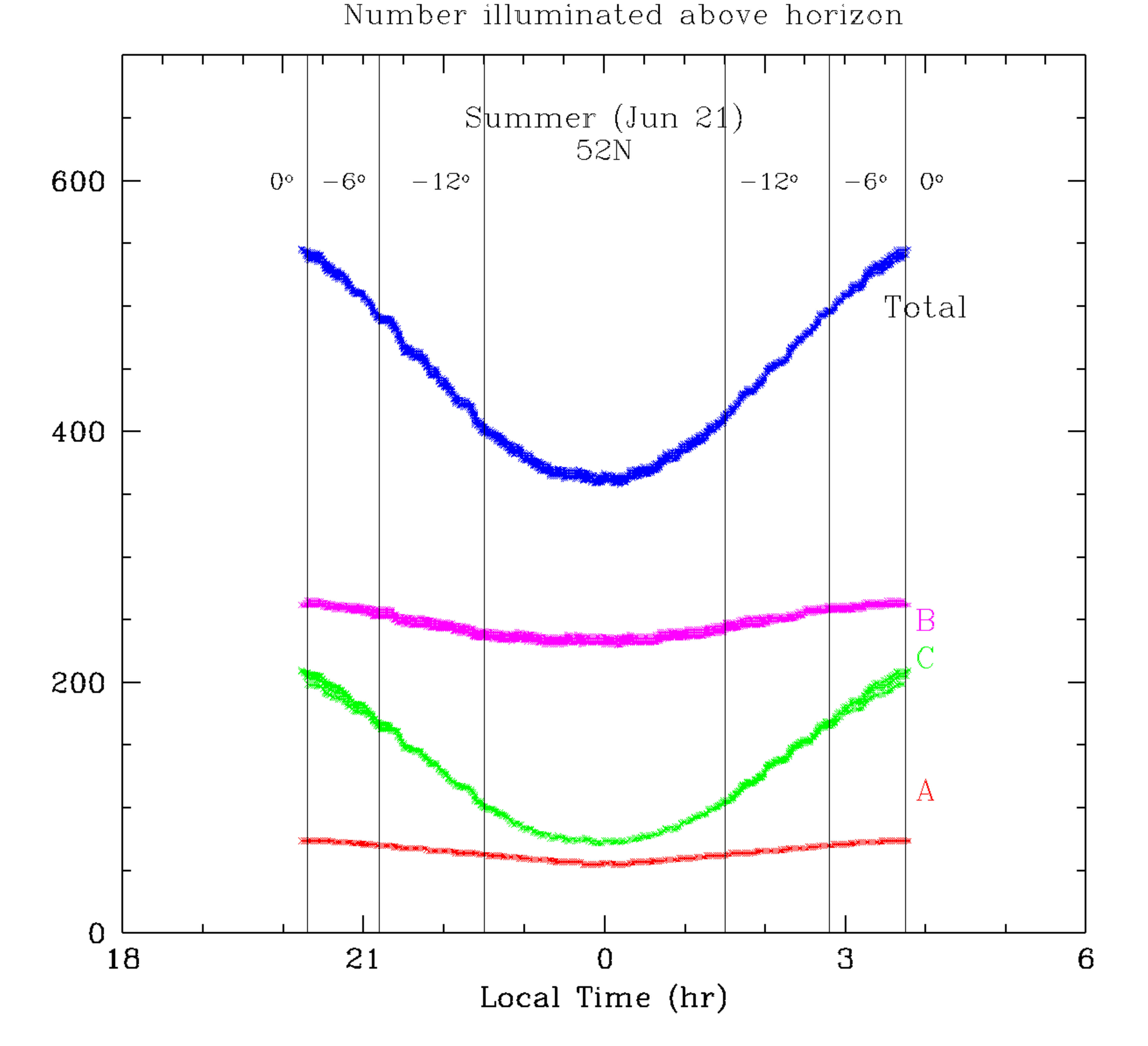}{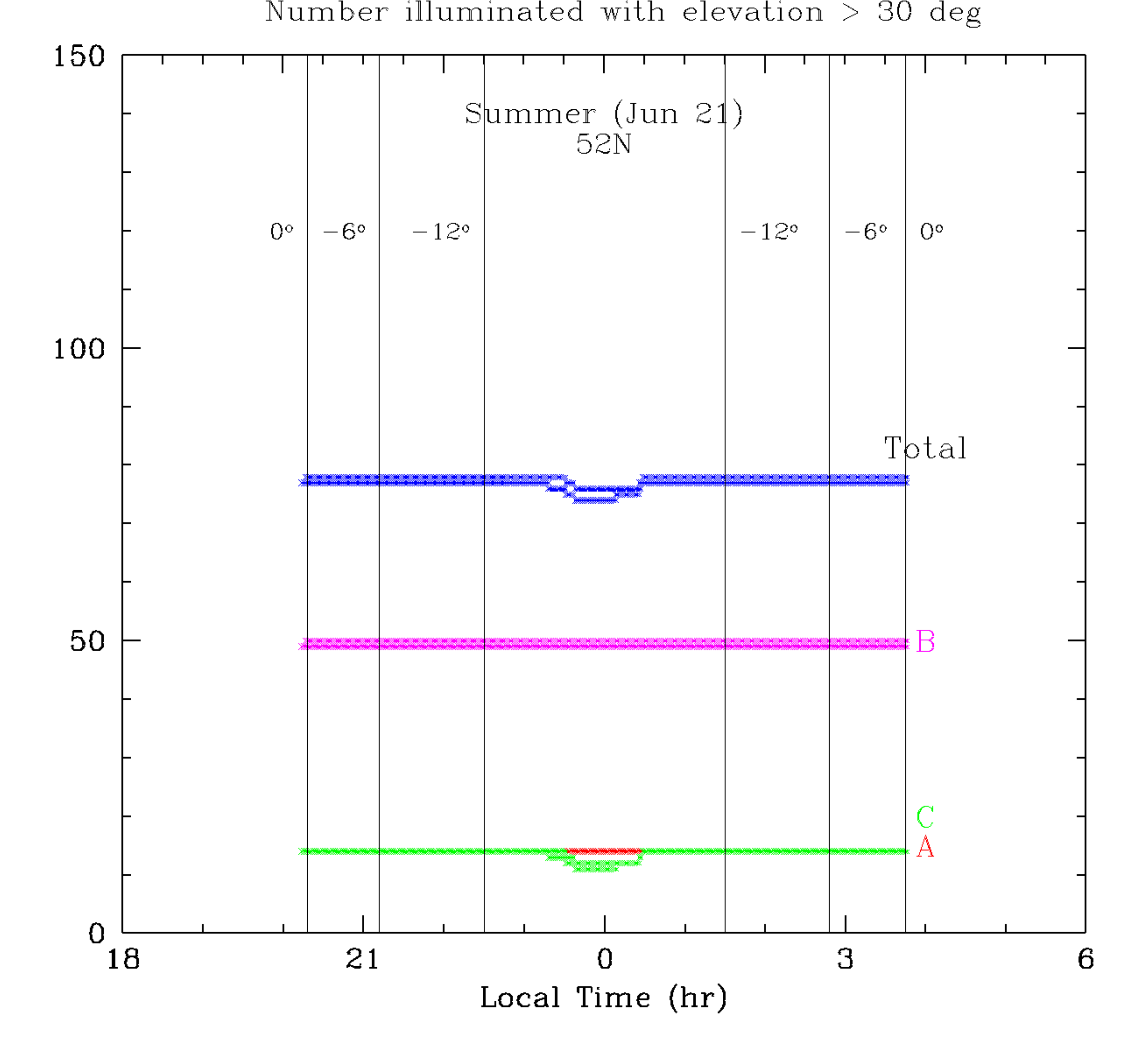}
\caption{Starlink satellites visible from London (52N) in summer, versus time of night.
Number above horizon (left); number above 30 deg elevation (right). {\rev Vertical lines indicate
the times at which the sun reaches elevations 0, -6, and -12 degrees for the ground observer.}
In these and subsequent figures, the separate contributions from layers A, B and C are 
shown as indicated.  The main contribution is from { \rev fainter Layer B satellites (V$\sim$7.5, see section \ref{sec:obs}).
Nevertheless, the model
suggests of the order of 25 bright layer A (V $\sim 5.5$) and C (V$\sim 4.5$) objects } visible at high
elevations at all times
through the summer night with a further 75 close to the horizon.
\label{fig:simu1}}.
\end{figure}

\begin{figure}[ht!]
\plottwo{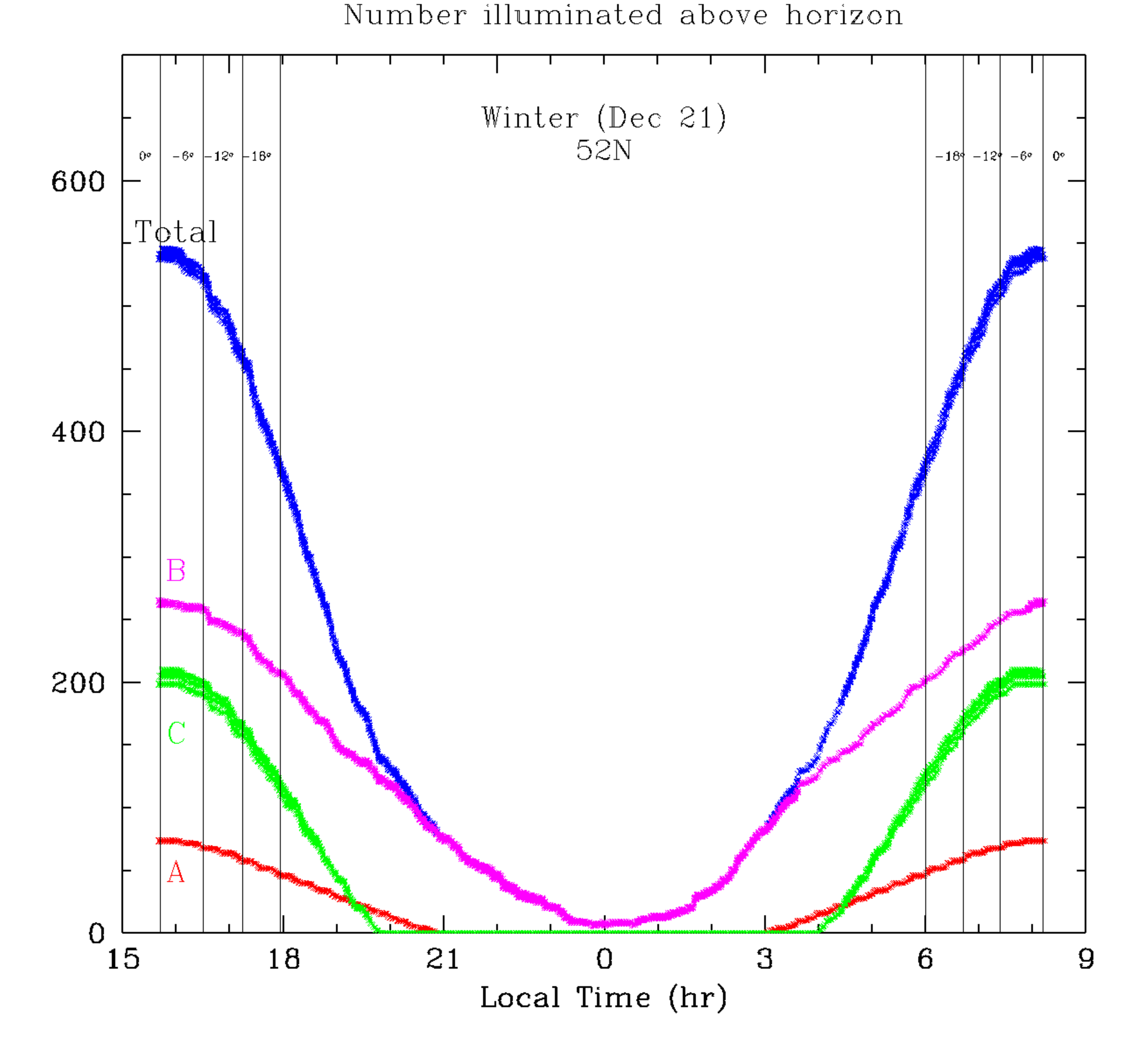}{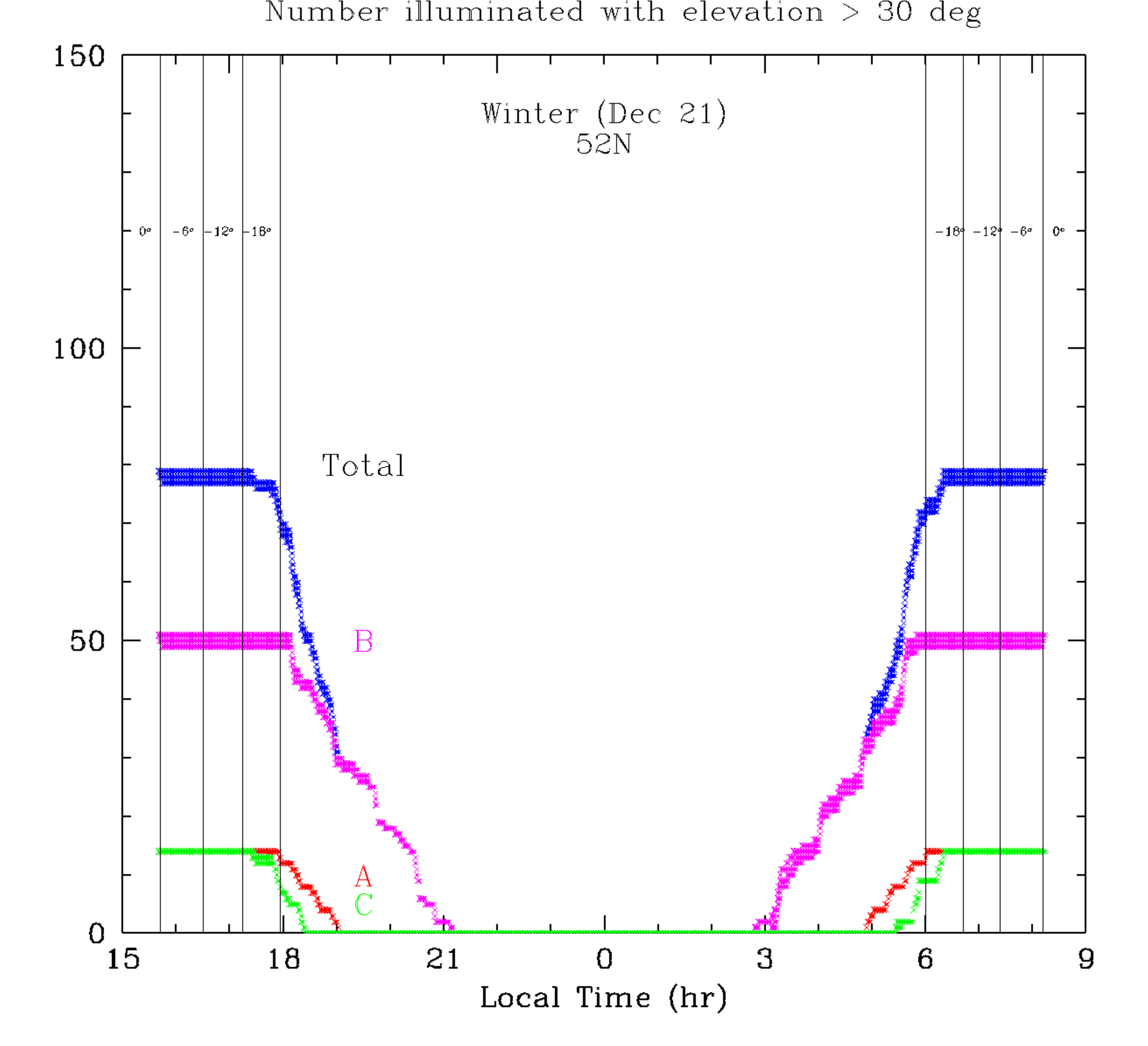}
\caption{Starlink satellites visible from London (52N) in winter, versus time of night.
Number above horizon (left); number above 30 deg elevation (right).
{\rev Vertical lines indicate
the times at which the sun reaches elevations 0, -6, -12 and -18 degrees for the ground observer.}
The sky should be free of naked-eye satellites in the middle of the night,
but there will up to 200 near the horizon during twlight.
\label{fig:simu2}}.
\end{figure}

\begin{figure}[ht!]
\plottwo{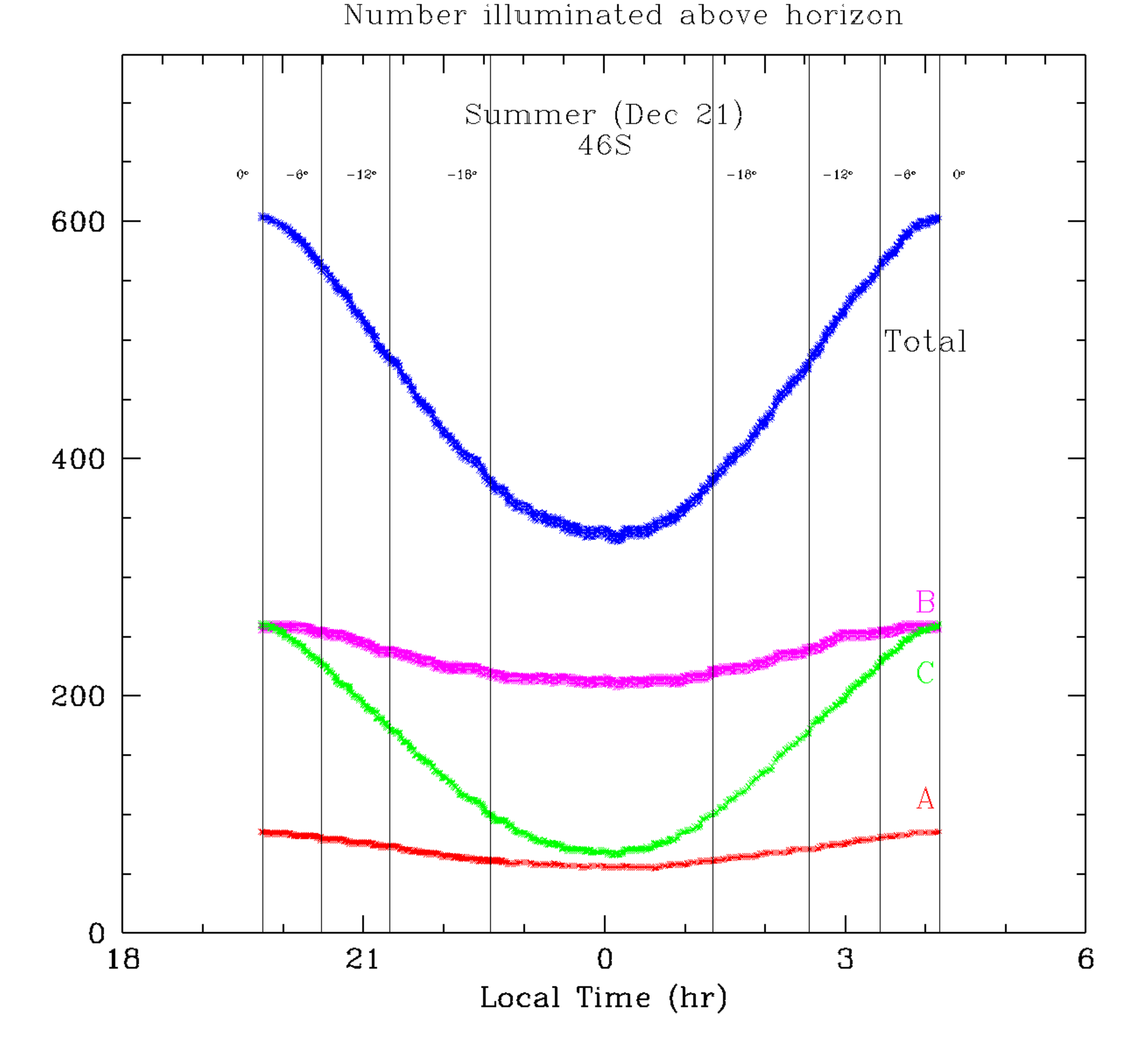}{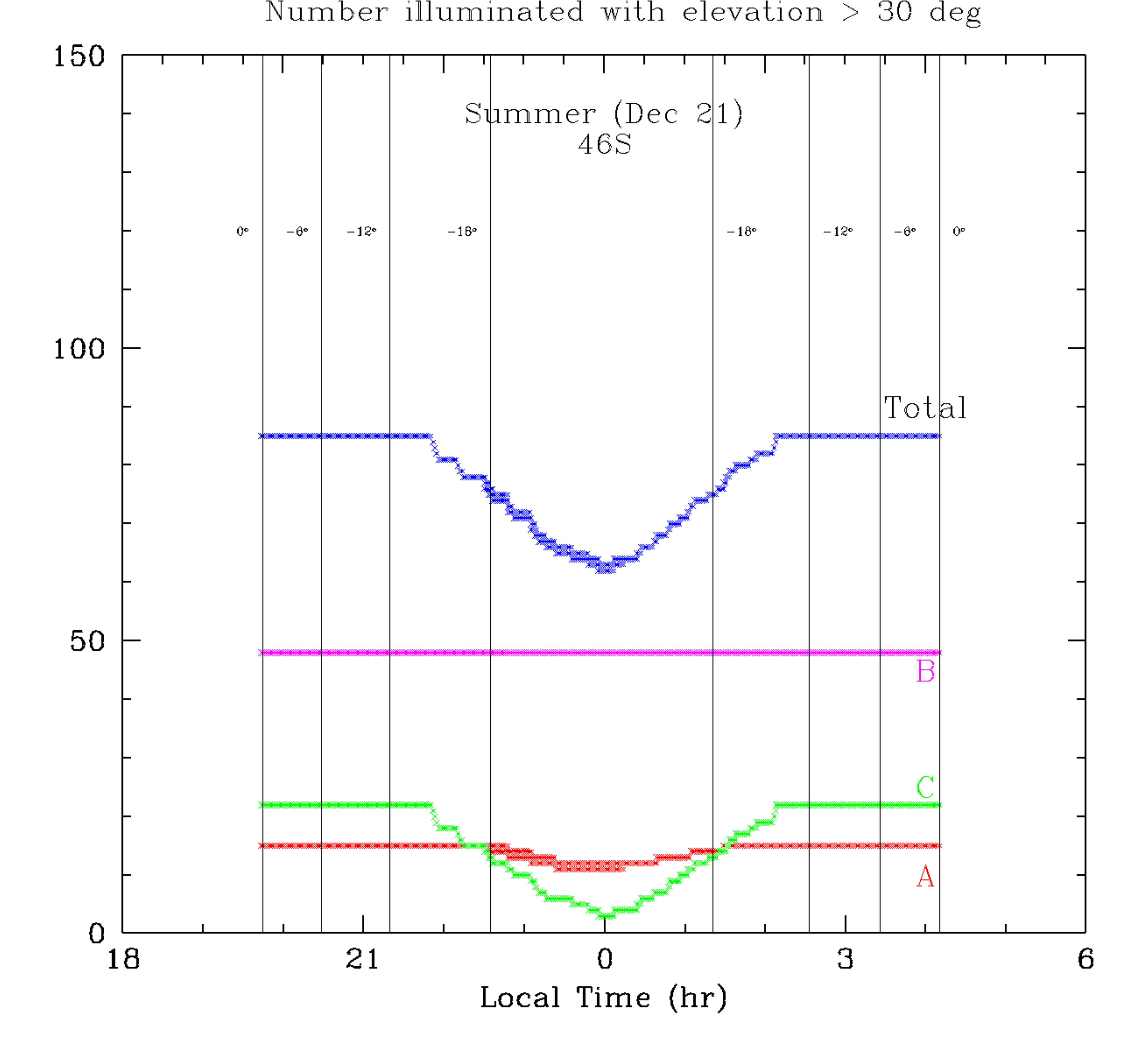}
\caption{Starlink satellites visible from Dunedin (46S) in summer (January), 
versus time of night.
Number above horizon (left); number above 30 deg elevation (right).
Results similar to latitude 52 deg, except that satellites in the lowest altitude
layer (C) are in shadow in the middle of the night.
  \label{fig:sim1}}.
\end{figure}

\begin{figure}[ht!]
\plottwo{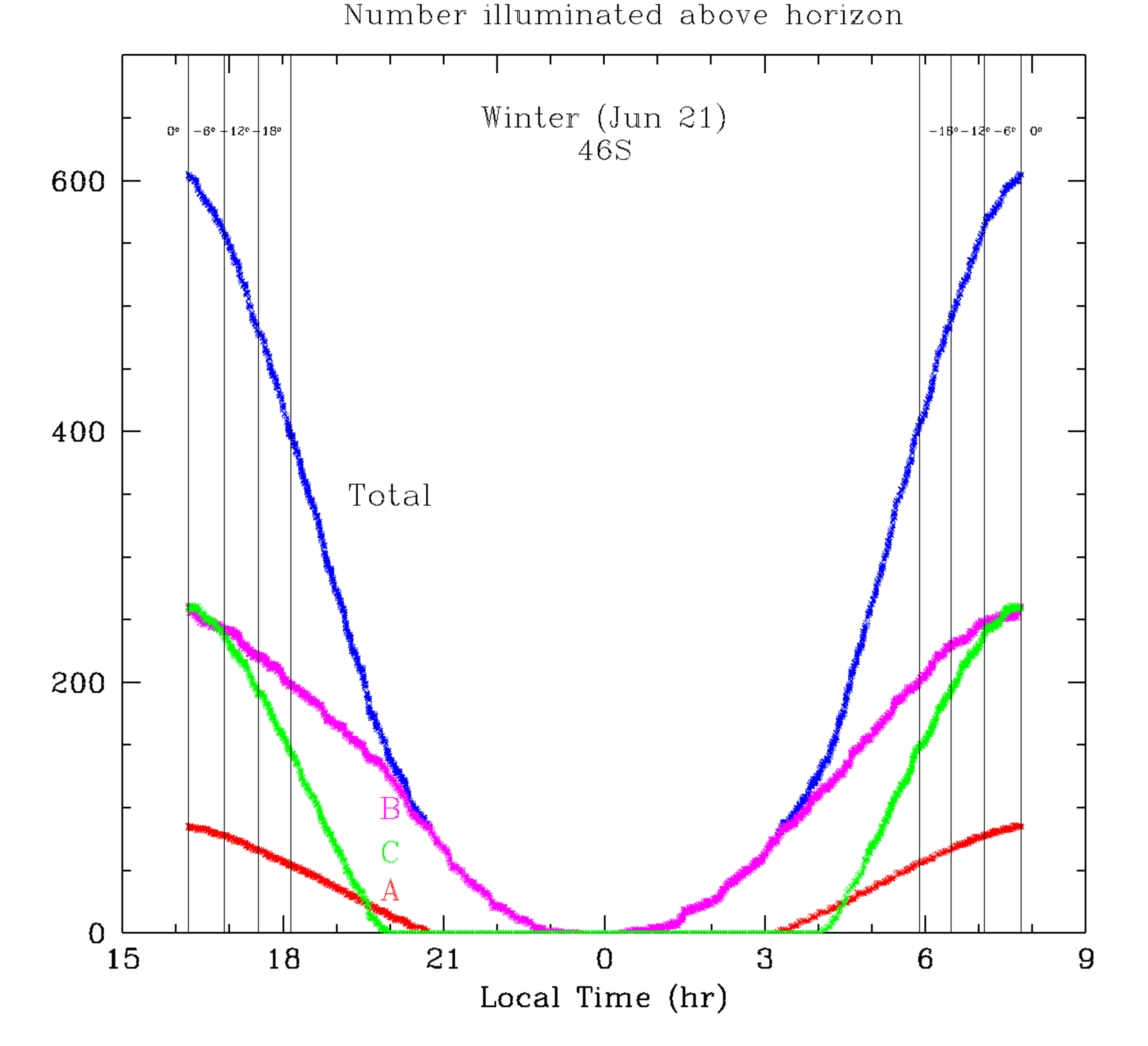}{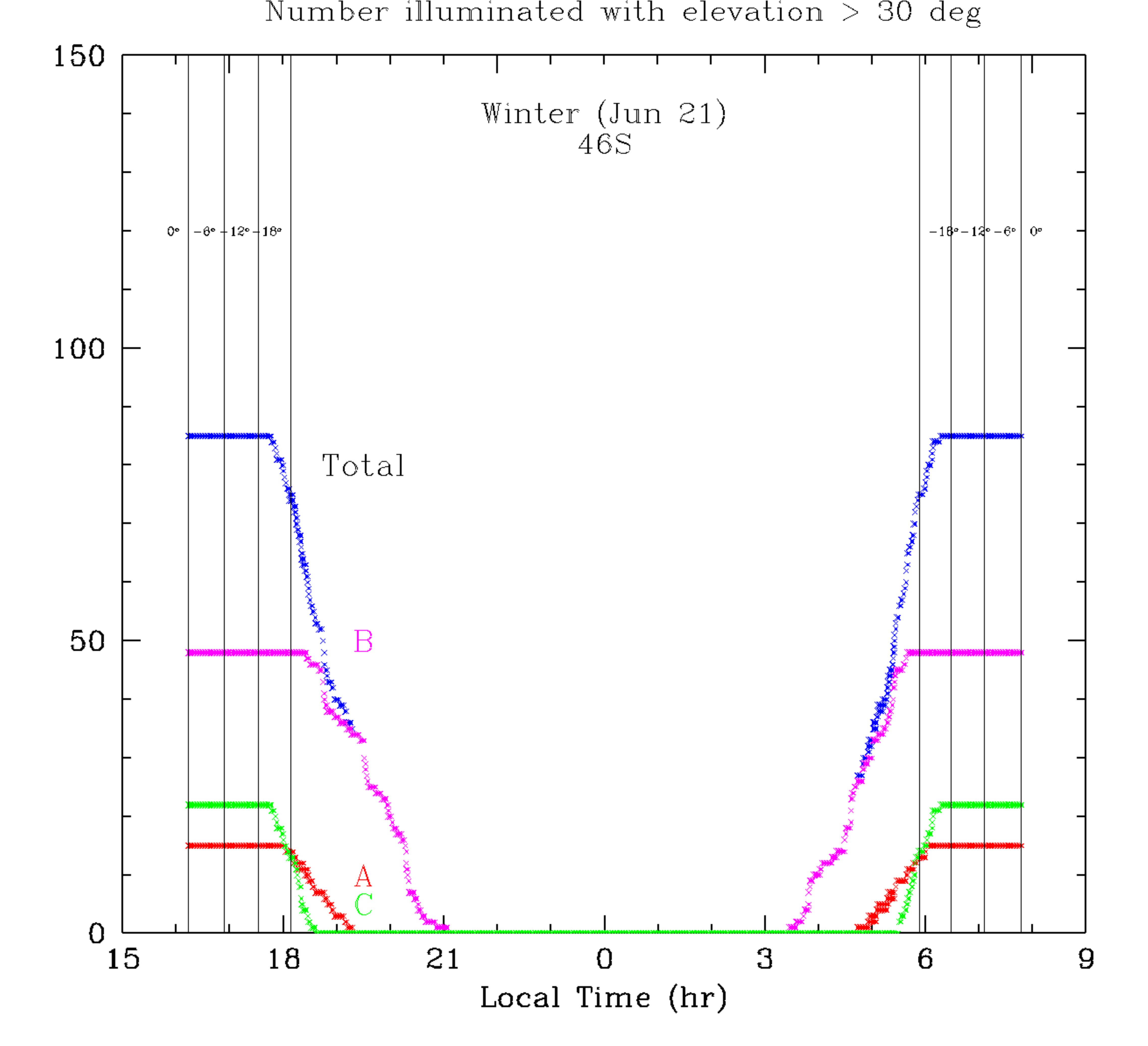}
\caption{Starlink satellites visible from Dunedin (46S) in winter, versus time of night.
Number above horizon (left); number above 30 deg elevation (right).
Satellites are only visible during twilight.
\label{fig:sim2}}.
\end{figure}

\begin{figure}[ht!]
\plottwo{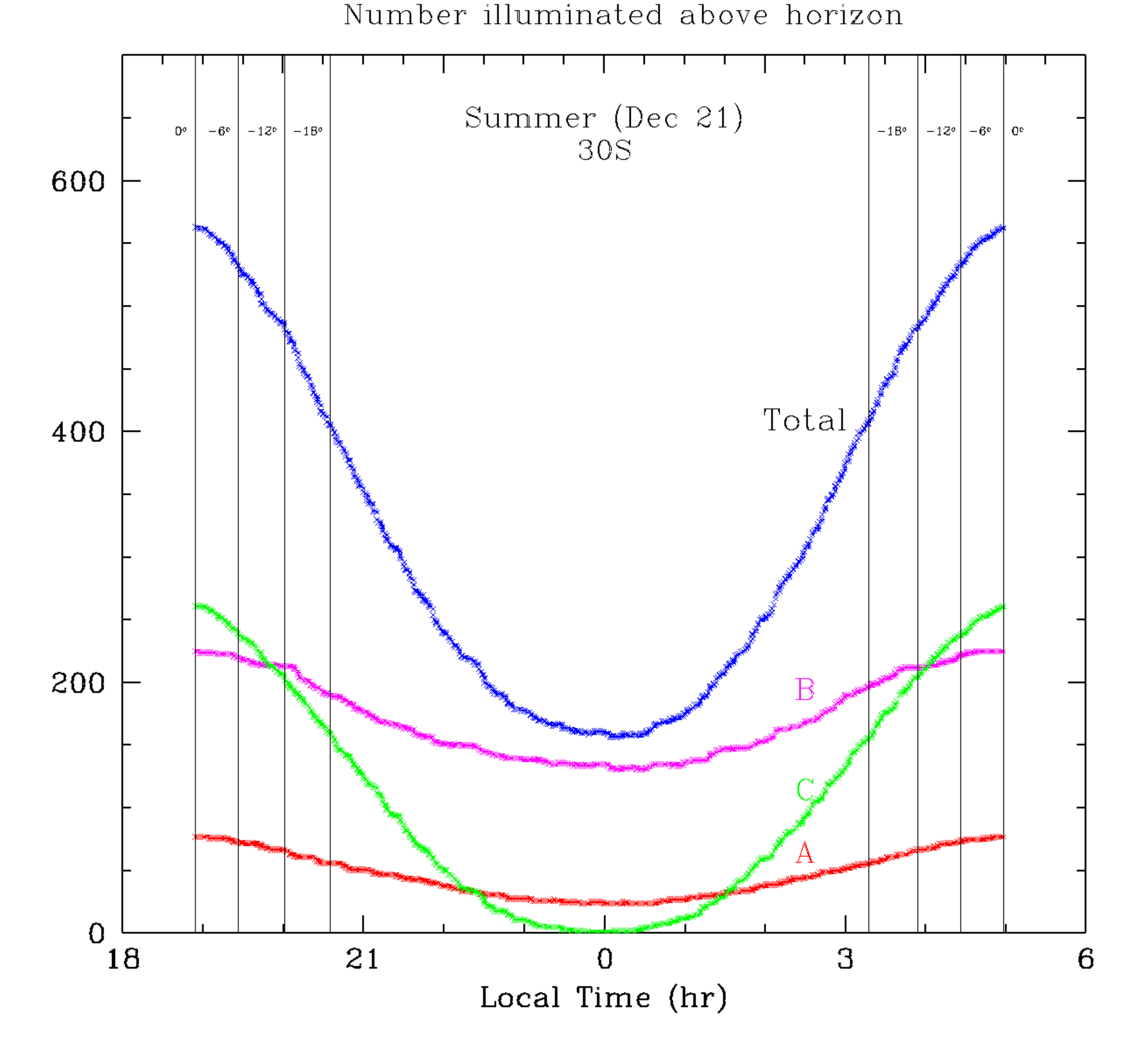}{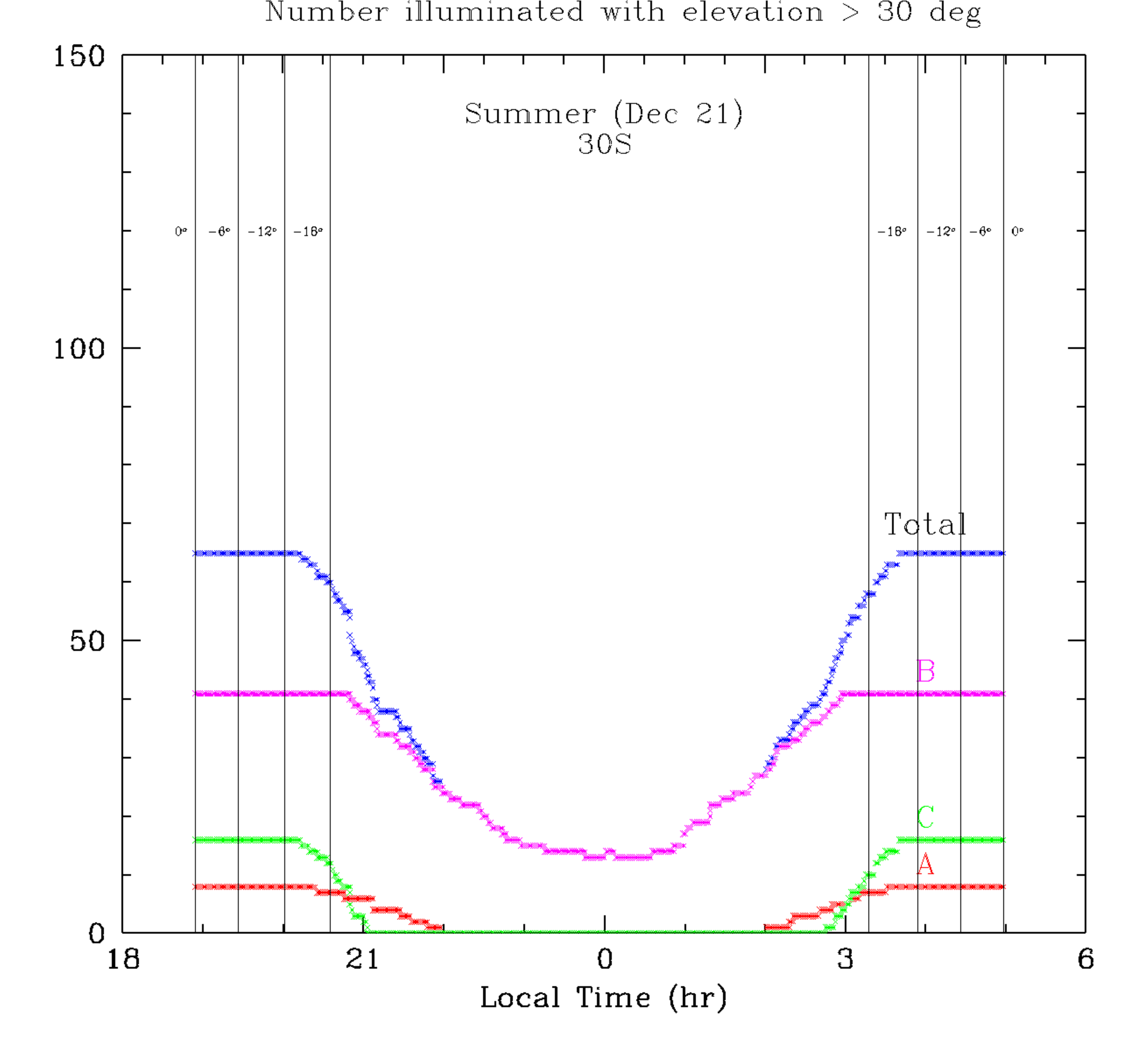}
\caption{Starlink satellites visible from Cerro Tololo (30S) in summer, versus time of night.
Number above horizon (left); number above 30 deg elevation (right) 
At this latitude, the naked-eye layers A and C are in shadow except at twilight.
At midnight 20 satellites are illuminated at high elevation corresponding
to $2\times 10^{-3}$ per sq degree.
\label{fig:simc1}}.
\end{figure}

\begin{figure}[ht!]
\plottwo{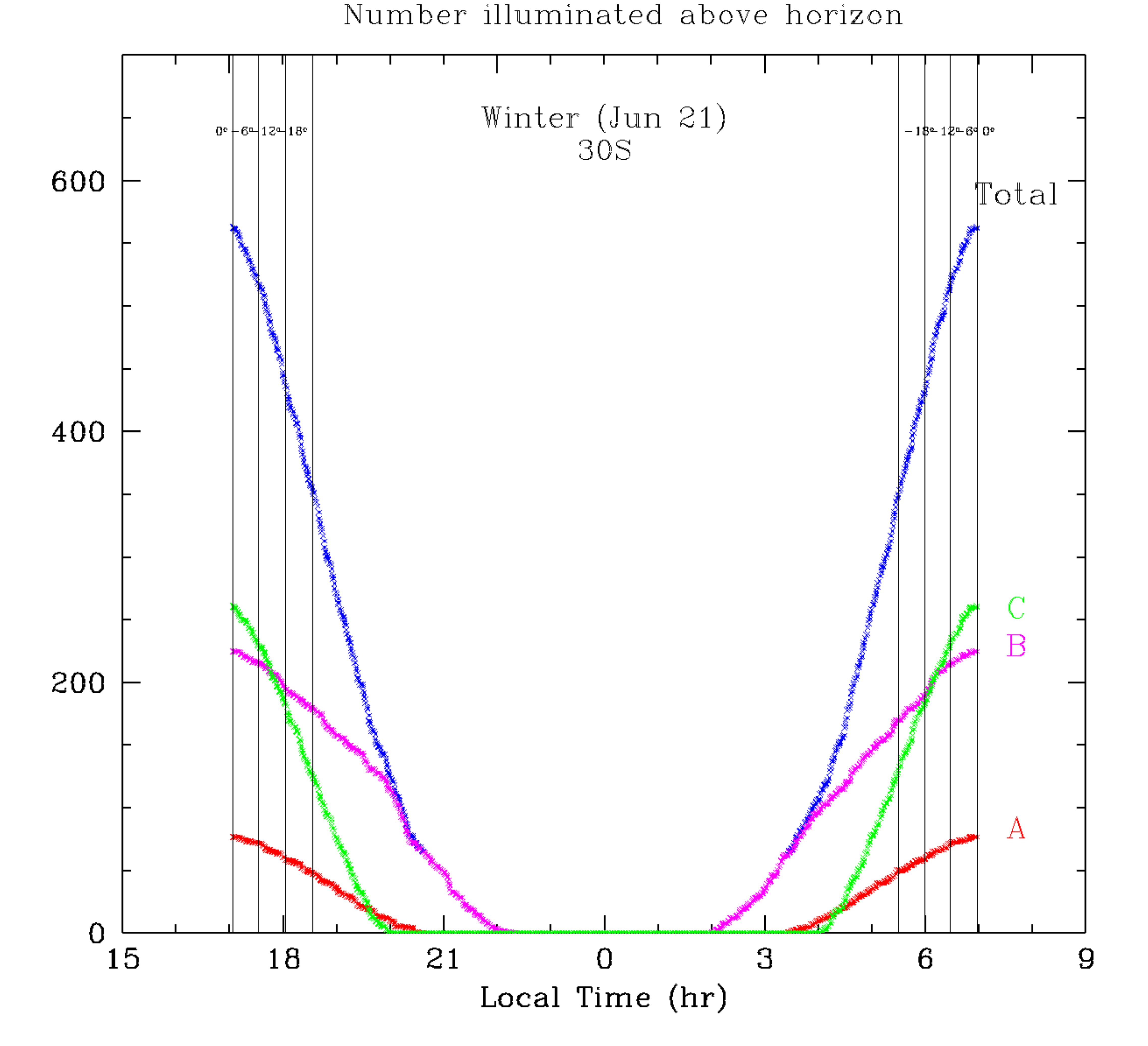}{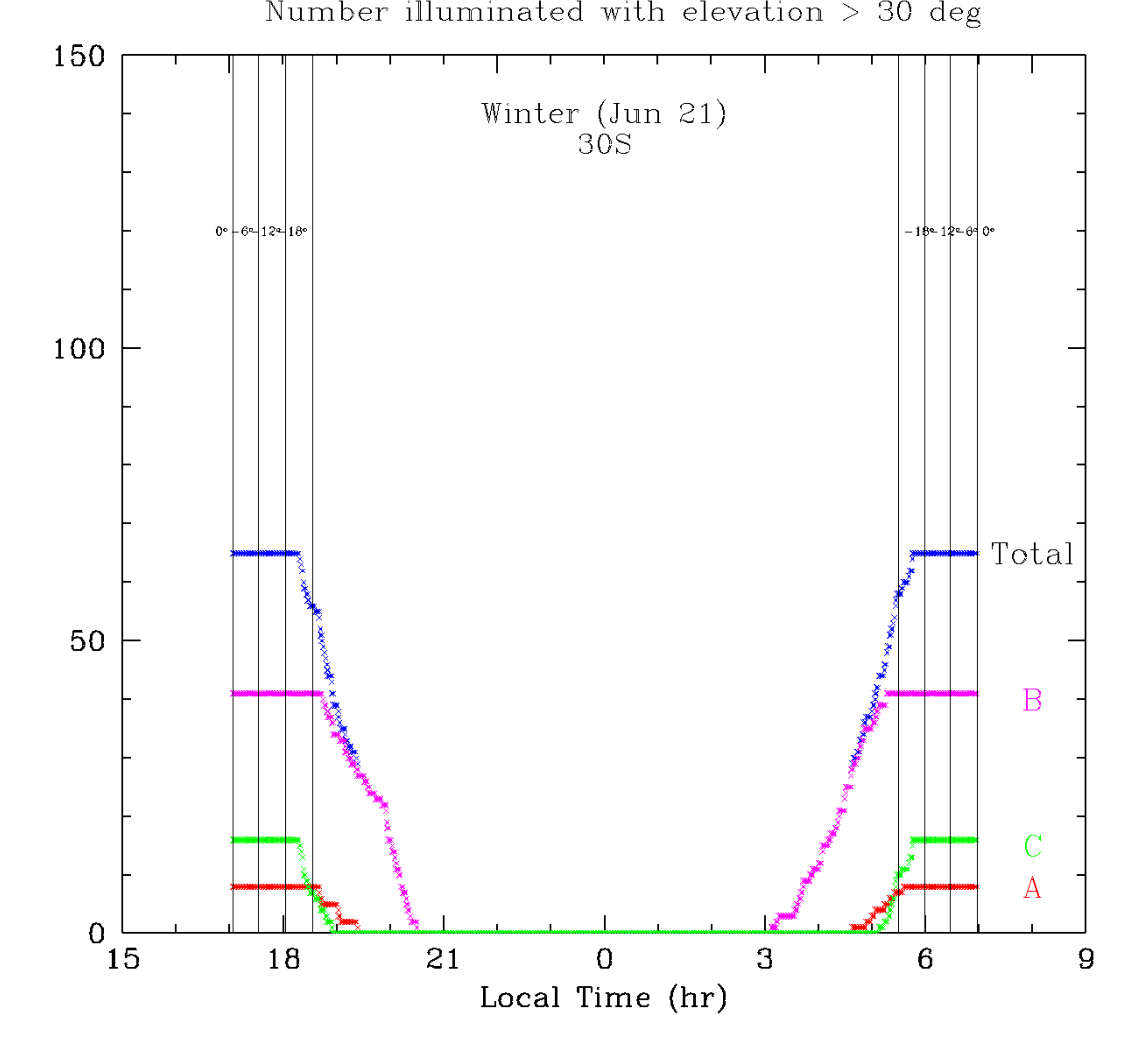}
\caption{Starlink satellites visible from Cerro Tololo (30S) in winter, versus time of night.
Number above horizon (left); number above 30 deg elevation (right).
Observations at twilight will be impacted, but about six hours
of unaffected time are available.
\label{fig:simc2}}.
\end{figure}

As representative
examples I evaluate summer and winter periods for three latitudes: a
typical populated Northern location at 52N
(London; figures \ref{fig:simu1}, \ref{fig:simu2}), 
a typical low density (so darker sky) Southern location at 46S (Dunedin, New Zealand; figures \ref{fig:sim1},\ref{fig:sim2}), 
and an astronomy-intensive location at 30S (Cerro Tololo, Chile; figures \ref{fig:simc1}, \ref{fig:simc2}). I consider both
the number of illuminated satellites above the horizon and the number illuminated above elevation 30 degrees (airmass 2),
representative of the impact to the general public and the impact to professional observations respectively.
We see that several hundred satellites are above the horizon at all times of night; during winter
twilight, and all summer night long, most of them are illuminated.
Note that 100 satellites with elevation above airmass 2 corresponds
to an average number density of about 0.01 per sq degree of sky.

\begin{figure}[ht!]
\plottwo{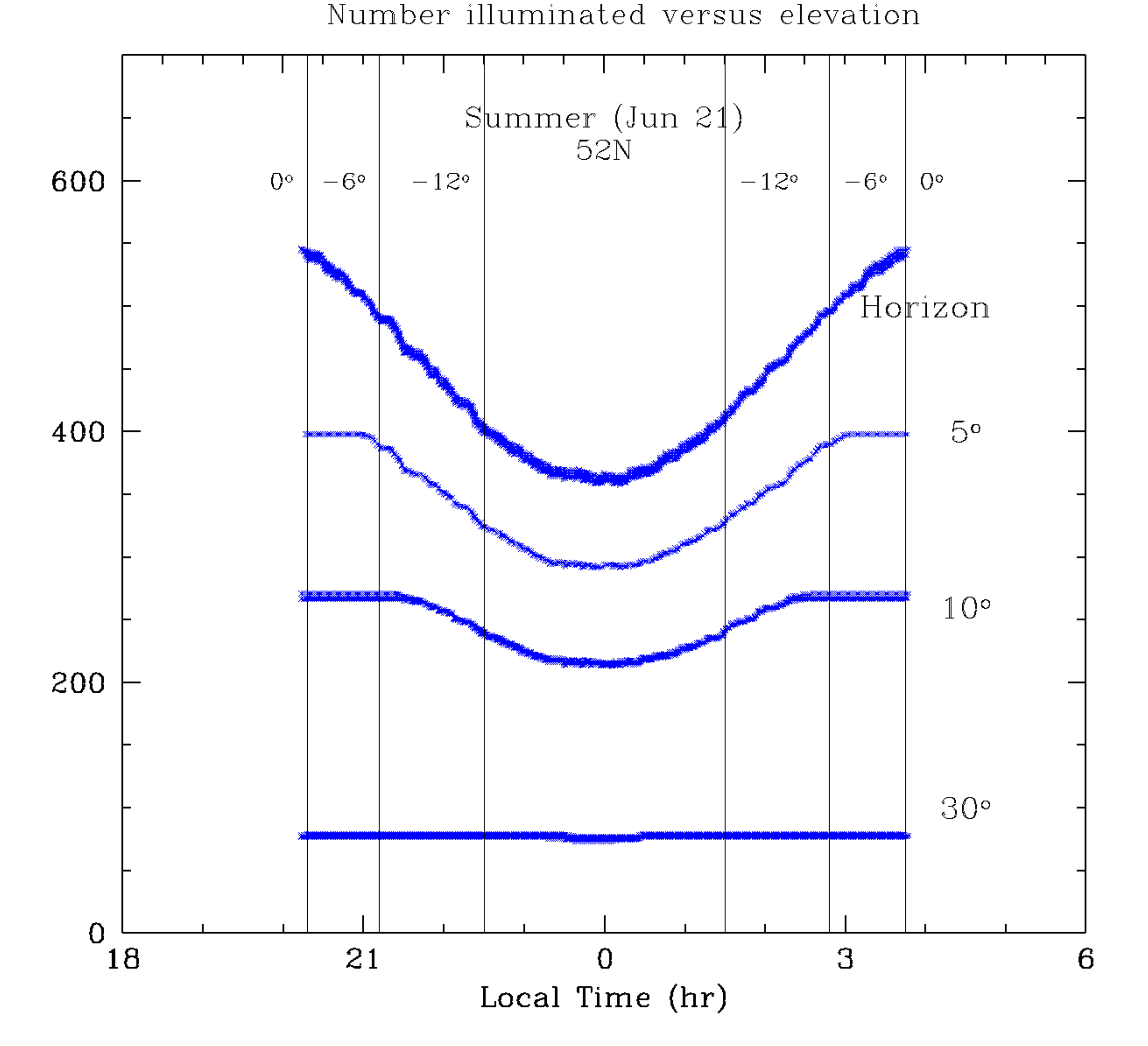}{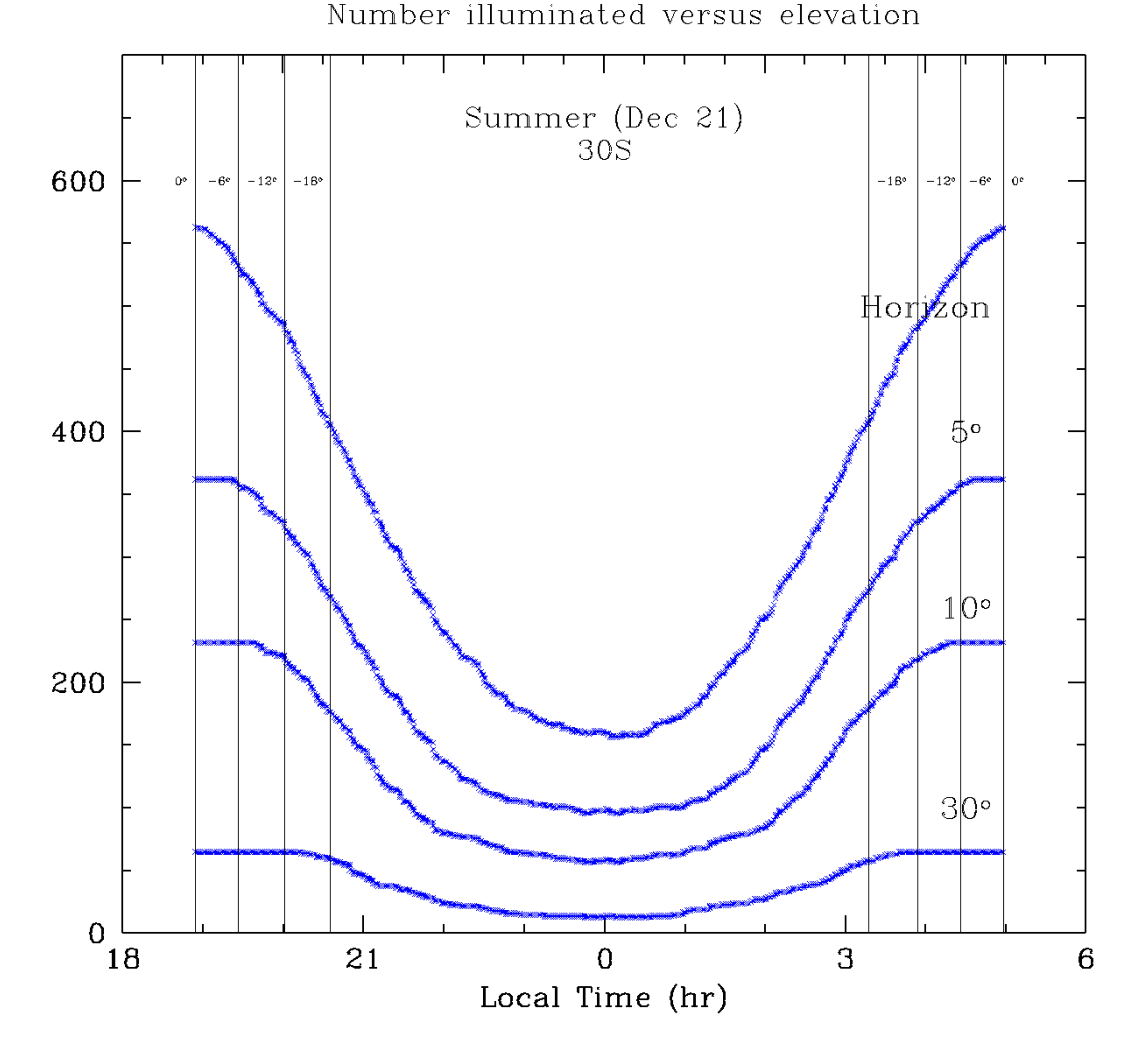}
\caption{\rev Starlink satellites (sum of layers A, B and C) visible from 52N and 30S in summer, versus time of night,
above a given elevation for elevations of 0, 5, 10 and 30 degrees. This shows that most of the 
illuminated satellites are near the horizon. but almost half have elevations greater than 10 degrees.
\label{fig:elev}}.
\end{figure}

{\rev The illuminated satellites are preferentially near the horizon. Figure \ref{fig:elev} repeats the content
for the total (sum of layers A to C) illuminated satellites fromf \ref{fig:simu1} and \ref{fig:simc1} for summer at 52N and 30S showing 
the numbers for elevations 0, 5, 10 and 30 degrees. Just under half the total are at elevations above 10 degrees (where they
will be visible above a typical cluttered and hazy horizon).}

The model neglects the effects of satellites in their orbit raising
phase. The current launch rate is about one batch of 60 per month and
the deployment strategy can be monitored using the publicly available
orbital data. Each launch places 60 satellites in orbit at around 300
km. Twenty of them undergo direct orbit raising to the 550 km altitude,
which takes 45 days. Two other groups of 20, however, are held at
350 km for 35 and 70 days respectively, using differential nodal
precession to  reach orbital planes separated by 20 and 40 degrees.  In
steady state this results in about 60 to 100 satellites in this lower
orbit where they are about 1 mag brighter than in the final orbit.  With
12,000 satellites each with a typical 5 year life\footnote{\label{Shotwell19}
Shotwell, G. cited in Sheetz, M., 
https://www.cnbc.com/2019/11/11/watch-spacex-livestream-launching-second-starlink-internet-mission.html},
the satellite replacement rate will rise to 200 per month.

\clearpage

% 10313

\clearpage

\section{Observations} \label{sec:obs}

\begin{figure}[ht!]
\includegraphics[height=5.0in]{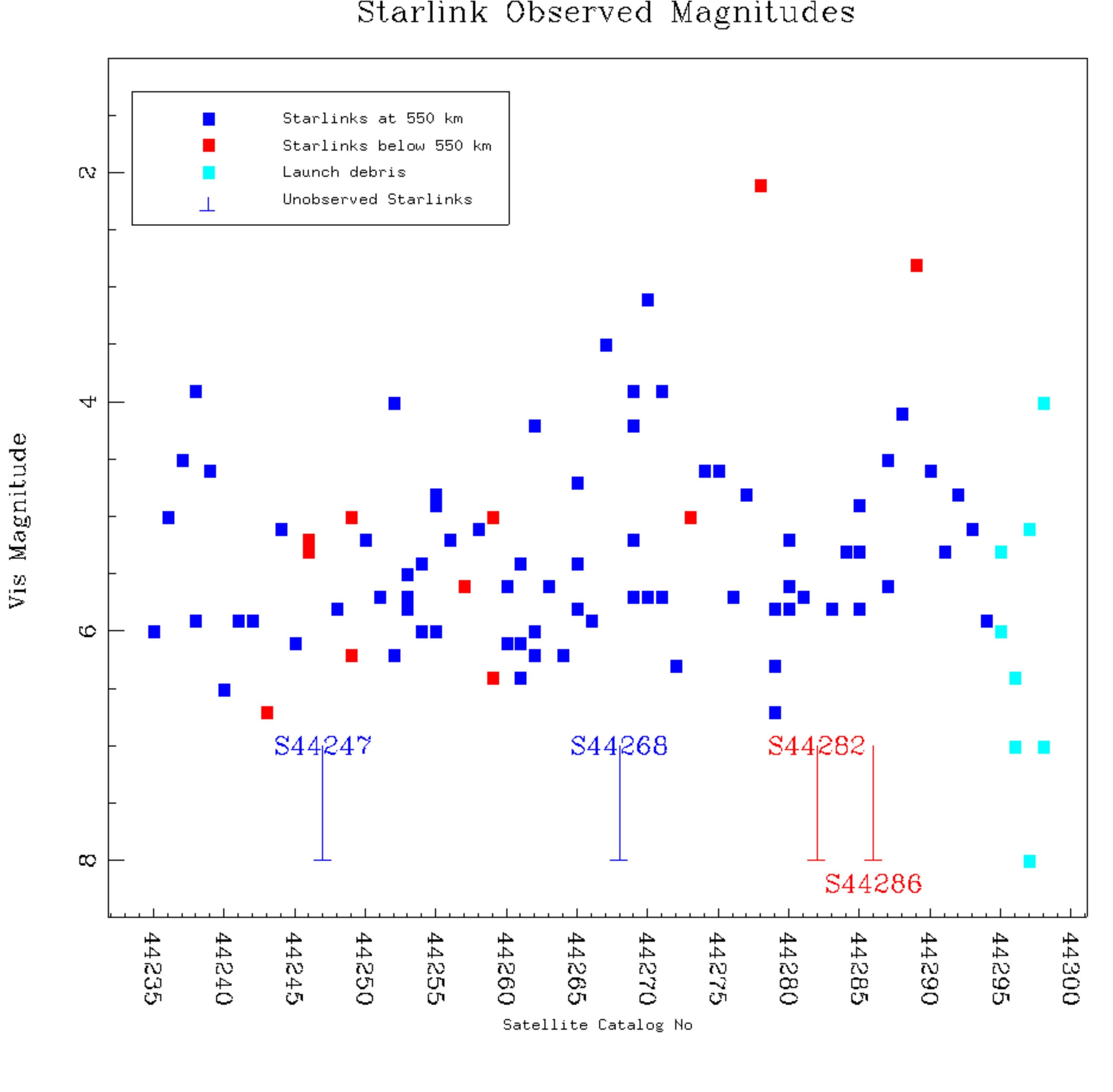}
\caption{Observed magnitude versus satellite catalog number for the 
first batch of Starlink satellites observed in summer 2019. Blue: satellites
in 550 km orbit. Red: satellites in lower orbits. Cyan: Deployment debris
objects. This shows that almost all the satellites from the first launch have
similar brightness; four of the sixty satellites were not seen and are indicated with their catalog numbers
at the bottom of the plot. \label{fig:v1}
}
\end{figure}

\begin{figure}[ht!]
\includegraphics[height=4.0in]{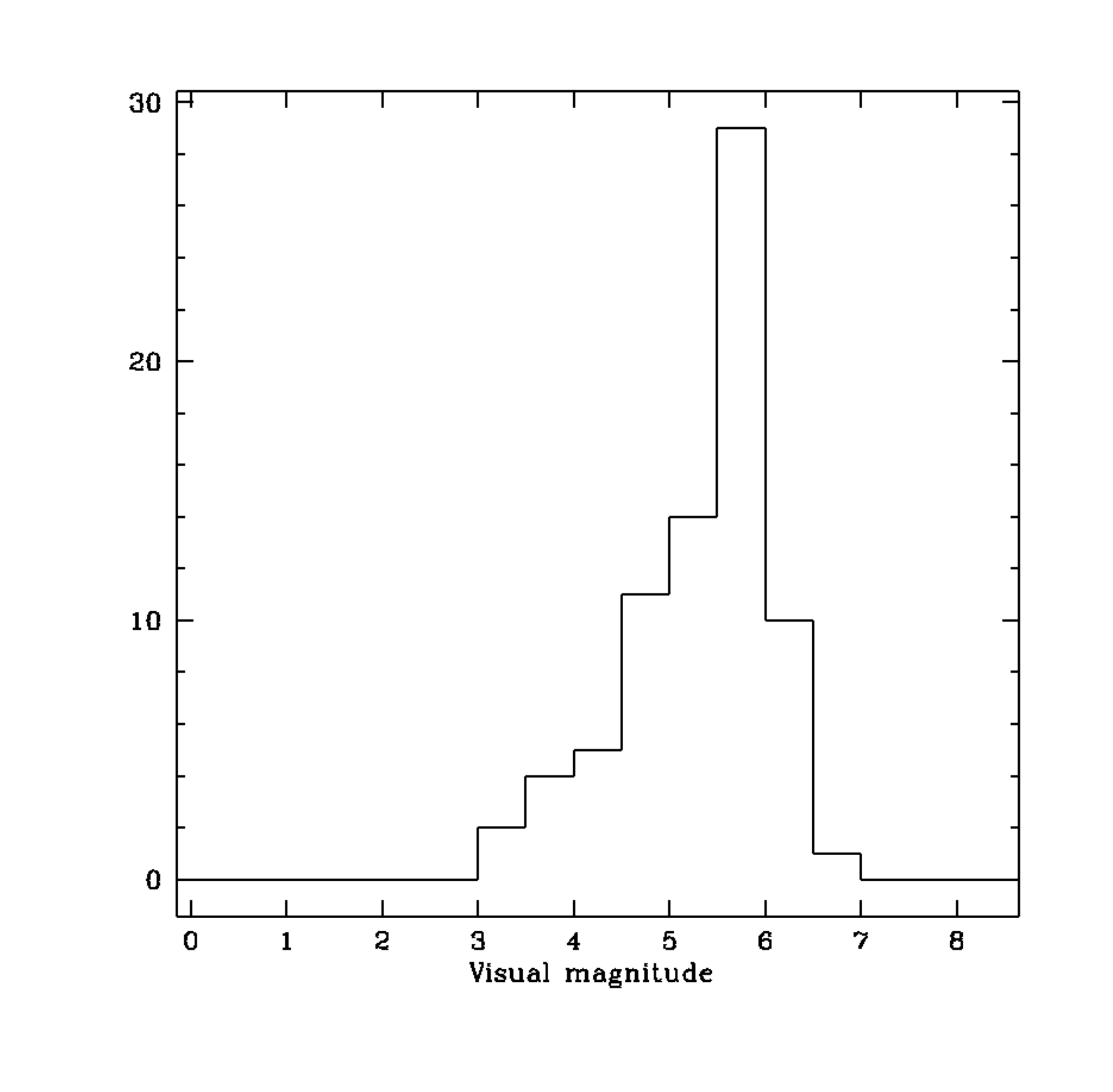}
\caption{Observed visual magnitude distribution for the 
first batch of 60 Starlink satellites observed in summer 2019 - satellites
in 550 km orbit only. \label{fig:v2}}
\end{figure}

The impact of the Starlink constellation depends on how bright the
satellites are found to be. Accordingly,
at the author's request experienced observers from 
the hobbyist satellite observing group SeeSat \footnote{\label{seesat}http://www.satobs.org} 
obtained visual observations
of almost all the first batch of Starlink satellites in the summer of 2019 (Figure \ref{fig:v1}). 
At this time the 
satellites were in the 550 km nominal orbit for the initial constellation (layer A), with
the exception of a few which had experienced anomalies. The satellites were observed
at a variety of elevations and phase angles, but a consistent picture emerges (Figure \ref{fig:v2}): the visual magnitudes
range from 3 to 7 with most between {\rev visual mag $5.5\pm 0.5$}. These observations confirm widespread press reports that
the majority of the satellites are naked-eye objects from dark sites 
most of the time when illuminated. {\rev
Detailed modelling of the phase-dependent brightness of the satellites versus elevation is beyond the
scope of the present paper. However, knowing the proposed altitudes, we can estimate the approximate zenithal brightness of future
satellites of the same design in layers B and C as $\sim 7.5 \pm 1.0$ and $\sim 4.5\pm 0.5$ respectively. Thus,
we expect the layer B satellites will not be naked-eye visible but the layer C satellites will be rather noticeable.
}

A new SeeSat observing campaign was begun in 2020 February, following
the launch of Starlink-1130 (`Darksat'), a non-operational satellite
whose nadir surface had a special coating. {\rev
Preliminary 
observations suggested that this coating had not been successful in
making this satellite fainter than the others. However, new results
in early 2020 March, after satellite had reached its operational
altitude and final orientation required to point the darkened face
toward nadir, indicate that it is now approximately 1 mag fainter that
other Starlink satellites at comparable altitude (\cite{Cole2020},
\cite{TregReed2020}). If similar modifications are made to future
satellites, the predictions made in this paper will need to be adjusted
accordingly.}

% projects/Sats/starlink sla

\clearpage
\section{Observational Impacts}

\cite{Seitzer20} and \cite{Tyson20} have discussed various impacts 
of bright satellites passing over the field of
view of an observation on
professional ground-based observations,
including image streaks, electronic crosstalk,
effects on flat fields, and ghost images. In addition, for
some telescopes a very bright satellite passing near the field of view
could add scattered light across the field, impacting the limiting
magnitude. Transient effects will occur even for non-illuminated
satellites, occulting celestial sources. However, these will be rare for
a given object and  the timescale for the occultation for a 1 arcsecond
seeing disk will be of the order of 20 milliseconds, so this should not be a
problem for most projects which are looking at much longer timescales.
 Streaks can also affect {\rev observations made from orbit by space telescopes}. When the 1100-1200 km layer B of the
Starlink constellation and the comparable OneWeb constellation are deployed, one can expect impacts to large 
(several square degree) field-of-view
observatories in lower orbits. With its narrow field of view, 
the Hubble Space Telescope (HST) should be less impacted,
although not immune - on 2020 Feb 28 an exposure was ruined by the pass of a Chinese rocket
stage only 34 km above the telescope, leaving a bright streak 23" wide across the target cluster
of galaxies (J. Schmidt 2020, personal communication).
\cite{Stankiewicz08}, \cite{Borncamp16}, and \cite{Borncamp19}
discuss the problem of identifying and masking satellite trails in HST/ACS images.

The Starlink satellites are also likely to be bright thermal infrared sources. I do not consider the
impact on near or mid-infrared astronomy in this paper, but further analysis seems warranted. Initial
estimates suggest they will be microJansky sources at cm wavelengths but very bright in the submm 
(Anita Richards, 2020, personal communication). 
The potential for radio inteference was, of course, anticipated before launch, and the radio astronomy
community has been working with SpaceX to minimize problems in that area.

The \cite{IAU19} and the \cite{AAS19} have issued statements of concern about the advent of the megaconstellations
and the AAS has established a working group on the subject \citep{AAS19b}. The working group has been
focussing on the impact to the Vera Rubin Observatory (formerly LSST) as representative of the worst impacts to observers. However,
there appear to be other science projects which may be more severely affected. For example,
searches for near-Earth asteroids include observations taken in twilight, a time when the satellites are 
illuminated year-round. Twilight surveys include the ZTF twilight survey \citep{Ye20}
and the proposed LSST Twilight Survey
\citep{Seaman18}.

Also, long exposures of wide fields are likely to be affected even in
dark time during summer.  In all cases, one can assume that the satellite crosses the field of view in a time short compared to the exposure
time. The expected number of satellite streaks crossing an image, assuming the satellites are essentially randomly
distributed, is proportional to the product of the field of view and the exposure time. By exposure time here, we
mean for a single frame, rather than the total accumulated exposure for a target object. One can median-filter 
sets of frames to remove streaks, as long as the number of streaks is low enough that no pixels are under a streak
in multiple frames. {\rev However, for many science projects there are overheads associated with each frame
that may make slicing the exposure into a larger number of shorter frames infeasible.}

If the angular surface number density of satellites (number per square
degree) is S, and their angular velocity is $\omega$ (degrees per
second), then the expectation value for the number of streaks across an
exposure of duration T (seconds) with field of view diameter D
is (to within geometrical factors of order 1)
\[
  N = S\omega X
\]
where $X = T * D$ characterizes the susceptibility of the observation to streaks. Note that this is basically
an extension of the classic Buffon's Needle problem \citep{Buffon1777}.

The terms due to the constellation vary depending on the orbits, time of year, etc. They can be scaled
to typical
`summer midnight' values of $ S\sim 100 \mbox{deg$^{-2}$}$ and angular velocity for 550 km orbit at zenith of 0.79 deg/s, and picking X=60 deg s:
\[
  N =  0.47 \left ( S \over 0.01 \mbox{deg$^{-2}$} \right ) \left ( \omega \over 0.79 \mbox{ deg s$^{-1}$} \right ) \left( T \over 60 \mbox{s} \right) \left( D \over 1 \mbox{deg} \right)
\]

We see that it is not implausible to reach $N\sim 1$, with every exposure having a satellite streak on average.
For the Vera Rubin Observatory, X is typically 15 s * 3 deg = 45 deg s. For CFHT/Megacam,
the Outer Solar System Origins Survey \citep{Bannister16}, \citep{Bannister18}
used a 1 sq deg FOV
with up to 400 s exposures, so X = 400 deg s. The TNO searches of \cite{Shep19} used CTIO/DECam (2.7 sq deg)
with 420 s exposures, so X  = 690 deg s. Thus, these examples are factors of ten more vulnerable to
satellite streaks than VRO. 

\section{Other Megaconstellations}

Other currently planned megaconstellations will have less impact on naked-eye observers since their proposed satellites
are both smaller and in higher orbits. However, comparable impacts on professional astronomy are likely. OneWeb Satellites
(https://onewebsatellites.com), 
based in the UK, is the other system currently being launched.
By 2020 March 1, 49 OneWeb satellites had been deployed; their planned operational orbit is around 1170 km with an
inclination of 88 degrees. No photometric observations of OneWeb satellites have yet
been reported. The OneWeb system can be compared with the simulations for layer B in my Starlink model, although the higher
inclination will result in a different latitude behaviour.
China has several less ambitious low orbit constellations planned, including Xinhe (1000 proposed satellites), Hongyun (864 proposed
satellites) and Hongyan (320 proposed satellties). Each of these projects has launched at least one test satellite to date.

\cite{OH20} have recently performed similar analyses for generic
constellations, with a somewhat different method. They derive broadly
compatible results, except  that they neglect the larger number of
visible satellites above 30 degrees elevation expected in northern
Europe and other high latitudes. {\rev  Additionally, we report observations that 
disprove their parenthetical suggestion that } current Starlink satellites are as
faint as mag 8 in their final orbits.

\section*{Conclusion}

The population of large artificial satellites in orbits below 600 km is
undergoing rapid change and is now dominated by  the Starlink system.
Starlink is the first of the megaconstellations to see significant
deployment, but it is unlikely to be the only one. Astronomers - and
casual viewers of the night sky - must expect a future in which the low
Earth orbit population includes tens of thousands of relatively large 
(few arcsecond angular size) satellites with a sky density of order 0.01
per square degree at zenith acting as sources of reflected sunlight
affecting ground-based (and in some cases even space-based)
observations. The impacts will be  significant for certain types of
observation (e.g. twlight observations and long-exposure observations
with wide fields of view), certain observatories (those at relatively
high latitude) and at certain times of year (local summer).

I am especially grateful to the SeeSat observers who provided Starlink magnitude estimates used here: Jay Respler, Brad Young, Cees Bassa, Bram Dorreman,
and Ron Lee; and to Michele Bannister and Martin Elvis for extensive comments.
I also thank Pat Seitzer, Patricia Cooper (SpaceX), and Dirk Petry for useful discussions
and to the anonymous referee for helpful comments which improved the paper.
Parts of this work were supported by the NASA Chandra X-ray Center,
which is operated by the Smithsonian Astrophysical Observatory for and on behalf of the National Aeronautics Space Administration under contract NAS8-03060.
\clearpage

\bibliography{ms}{}
\bibliographystyle{aasjournal}

\end{document}